# On Jaky Constant of Oedometers, Rowe's Relation
# and Incremental Modelling

## P. Evesque

Lab MSSMat,  UMR 8579 CNRS, Ecole Centrale Paris
92295 CHATENAY-MALABRY, France, e-mail evesque@mssmat.ecp.fr

### Abstract:

*It is recalled that stress-strain incremental modelling is a common feature of most theoretical description of the mechanical behaviour of granular material. An other commonly accepted characteristics of the mechanical behaviour of granular material is the Rowe's relation which links the dilatancy* $K = -\partial\varepsilon_v/\partial\varepsilon_1$ *to the stress ratio* $\sigma_1/\sigma_3$ *during a* $\sigma_2 = \sigma_3 = c^{ste}$ *test , i.e.* $\sigma_1/\sigma_3 = (1+M)(1+K)$. *We combine these two features and extract an incremental pseudo-Poisson coefficient which varies with* $\sigma_1/\sigma_3$ . *We solve the oedometric-test case, starting from isotropic sample and stress, for which* $\sigma_1$ *is increased continuously. It is found that the* $\sigma_1/\sigma_3$ *ratio evolves towards an asymptotic value* $k_o$ *which depends on the friction angle* $\varphi$ *only. It is shown that this asymptotic value* $k_o$ *compares well with the experimental fit known as the Jaky constant i.e.* $k_{Jaky} = 1 - \sin\varphi$ , *where* $\varphi$ *is the friction angle.*

______________________________________________________________________________

The aim of this paper is to point out few simple experimental features, mainly the Rowe's equation and the Jaky law,  which are well accepted by the scientific community but which are not integrated in a single scheme and to try and propose such a unifying approach which links them together. This will be done using hypoelasticity.

***Rowe's law of dilatancy:*** Rowe (1962) has proposed some relationship which relates the stress field ($\sigma_1$, $\sigma_2 = \sigma_3$) in an axi-symmetric triaxial test and the dilatancy K of the material; K is defined as $K = -\delta\varepsilon_v/\delta\varepsilon_1$, where $\varepsilon_v$ is the volume deformation and $\varepsilon_1$ is the axial one. His approach consists in considering regular arrays of cylinders and in analysing the stress field which is required to impose the sliding of a given row of cylinders, taking into account the stress field and the friction at contact points. As the orientation of the row is in general not parallel to the local surfaces of contact, it results that i) the stress field required to impose the row motion does not correspond to that one of the sliding along the row direction, but does correspond to a sliding parallel to the surface of contacts, and that ii) the misfit of orientation between the row direction and the contact surfaces leads to a volume change. As these two mechanisms are linked together they impose some equation which links them together; this is the Rowe's equation.   It turns out that this relation can be written in such a way that it does not depend on the nature of the 2d lattice (triangular, square,…): $\sigma_1/\sigma_3 = (1+K)\tan^2(\pi/4 + \varphi/2)$ with $\varphi$ being the friction angle. So, Rowe has generalised his results and concluded that this relation should be valid whatever the lattice, even when it is 3d and disordered. Experimental tests have been performed to check the validity of the relation using axi-symmetric triaxial test at $\sigma_2 = \sigma_3 = c^{ste}$ and relatively good agreement has been found (Rowe 1962, Frossard 1983).

***Oedemetric test***: There is also another troublesome fact; it concerns oedometric experiment for which one measures the evolution of the stress ratio $k_o = \sigma_2/\sigma_1 = \sigma_3/\sigma_1$ imposed to a material submitted to an axi-symmetric loading when its radius is kept constant. Experimental data show that this stress ratio tends to a limit value which depends on the friction angle $\varphi$; it is known as the Jaky formula ($k_o = 1 - \sin(\varphi)$) (Jaky 1944, 1948, Terzaghi 1965, Tsechebotariov 1951), but there is no correct theoretical demonstration of this correlation to the best of my knowledge (Tsechebotariov 1951). For instance, this limit does not correspond to the plastic condition of a friction material ($k_{plastic} = tg^2(\pi/4 \pm \phi/2)$. It seems that the simplest approach to explain this experimental fact is to take into account the elastic response via the Poisson coefficient $\nu$ ; in this case elasticity yields to $k_{elastic} = \sigma_2/\sigma_1 = \sigma_3/\sigma_1 = \nu/(1-\nu)$. However, the existence of the correlation between the Poisson coefficient and the friction angle remains mysterious.

A question arises: put together, are all these results compatible or not? Can we describe them in a same approach? To discuss this point and to propose an alternative approach is the aim of this paper. This is done by using an incremental modelling. So, in a first step, we recall the basis of this approach and we recall that it is able to describe plastic mechanisms. In the second step, Rowe's relation is combined to an incremental modelling to find the stress ratio $k_o = \sigma_2/\sigma_1 = \sigma_3/\sigma_1$ in an oedometric test; this leads to a relation which is not the Jaky's formula but which is quite narrow it.

***Hypoelastic modelling vs. plastic modelling:*** Hence, let us restate the problem in the following way: let us first consider a sand sample submitted to a given stress field ($\sigma_1$, $\sigma_2 = \sigma_3$) and let us consider any incremental deformation $\underline{\delta\varepsilon} = (\delta\varepsilon_1, \delta\varepsilon_2, \delta\varepsilon_3)$; it is most likely that one can force the sample to deform according to this path by applying an increment of stress $\underline{\delta\sigma} = (\delta\sigma_1, \delta\sigma_2, \delta\sigma_3)$; hence, any set of infinitely small deformation $(\delta\varepsilon_1, \delta\varepsilon_2, \delta\varepsilon_3)$ is possible; in other words, the evolution of this sample is governed by an incremental law which relates the increment of stress tensor $\underline{\delta\sigma}$ to the increment of strain tensor $\underline{\delta\varepsilon}$ so that one can write a relation of the kind $g(\underline{\delta\varepsilon}, \underline{\delta\sigma}, \underline{\sigma}) = 0$; furthermore, as the evolution of a sand sample does not depend only on its present stress field but also on its story, g shall be a function which is story dependent. g is valid at first order in $\underline{\delta\sigma}$ and in $\underline{\delta\varepsilon}$ , but this remark is not so important since the response is already non-linear due to the fact that g contains the variable $\underline{\sigma}$.

*Hypoelasticity for granular materials:* Now let us introduce the objectivity principle, which states that the response $\underline{\delta\varepsilon}$ shall be unique for a given increment $\underline{\delta\sigma}$ applied to a given sample under specified condition; so one shall be able to write $\underline{\delta\varepsilon}$ in the form $\underline{\delta\varepsilon}$

=$f(\underline{\delta\sigma},\underline{\sigma})$, where f is a function which depends on the sample story. Moreover, owing to the existence of the quasi-static regime which states that the response $\underline{\delta\epsilon}$ of the material is independent of the speed of loading if this one is slow enough, it can be shown that f shall be a homogeneous function of degree 1 in $\underline{\delta\sigma}$ (i.e. $f(\lambda\,\underline{\delta\sigma},\underline{\sigma})=\lambda f(\underline{\delta\sigma},\underline{\sigma})$) (Darve 1987) ; this means that the response to an increment of stress $\underline{\delta\sigma}$ in a given direction shall vary linearly with its module $||\delta\sigma||$ . However, this incremental law can not be strictly linear as a function of $\underline{\delta\sigma}$ in the whole domain of possible increments $\underline{\delta\sigma}$, because such an hypothesis would generate perfect reversibility and would be in contradiction with the well known fact that the evolution of a granular material is not reversible: in particular, labelling $\underline{\delta\epsilon}$ the response to the increment of stress $\underline{\delta\sigma}$ , and $\underline{\delta\epsilon'}$ the deformation corresponding to $-\underline{\delta\sigma}$ one shall not have $\underline{\delta\epsilon'} = -\underline{\delta\epsilon}$ but shall have $\underline{\delta\epsilon'} \neq -\underline{\delta\epsilon}$ (Darve 1987) . This is why hypo-elastic law is commonly used to model the rheological behaviour of this medium (Darve 1987 ,Tejchman 1997). One of the simplest modelling consists in separating the space E of variation of $\underline{\delta\sigma}$ into few separate subspaces $E_k$ (with $E = \cup E_k$ ) where the rheological law $\underline{\delta\epsilon}=f_k(\underline{\delta\sigma},\underline{\sigma})$ is strictly linear within each domain in the limit of a $||\underline{\delta\sigma}||$ infinitely small. As it is currently observed experimentally that the response to a strain increment is continuous; we will assume such a property here too, which imposes some relation between the $f_k$ : the responses $f_k(\underline{\delta\sigma},\underline{\sigma})$ and $f_{k+1}(\underline{\delta\sigma},\underline{\sigma})$ at the frontier between the two zones $E_k$ and $E_{k+1}$ shall be equal, i.e. $f_k(\underline{\delta\sigma},\underline{\sigma}) = f_{k+1}(\underline{\delta\sigma},\underline{\sigma})$. It is worth noting that this continuity is achieved spontaneously for two opposite directions $\underline{\delta\sigma}$ and $-\underline{\delta\sigma}$, since the crossing occurs at $\underline{\delta\sigma}=0$ and due to the fact that f is homogeneous of degree 1 in $\underline{\delta\sigma}$..

*Hypoelasticity applied to plastic behaviour:* It has been demonstrated (Loret, 1985 a & b) that this incremental approach is able to describe systems obeying perfect plasticity theory and/or elasto-plastic one with one or few different plastic mechanisms. In the case of an elasto-plastic system with a single plastic mechanism, the direction of the plastic deformation of the sample is controlled by the direction of the normal to the load surface, and the amplitude of the plastic deformation is controlled by the hardening law so that it depends linearly on $||\underline{\delta\sigma}||$; the total (elastic+plastic) deformation is then the sum of an elastic response and of a plastic yielding in a precise direction, both being proportional to $||\underline{\delta\sigma}||$ . Since projection operators act as linearly independent mechanism and because they can be added linearly, it turns out that a sample obeying to an elasto-plasticity theory with multiple-mechanism law, all being activated, shall obey the incremental modelling with a linear response by zones in the limit of a large-but-finite number of independent plastic mechanisms.

In the rest of the paper, a granular material will be assumed to obey such an incremental description with a set of linear functions $f_k$ , each defined for a zone; furthermore, the number of zones will be assumed small enough, so that triaxial tests performed at $\sigma_2 = \sigma_3 = c^{ste}$ and at constant mean stress p (i.e. $3p = \sigma_{10} + \sigma_{20} + \sigma_{30} = c^{ste}$) pertain to the same linear domain.

We will consider the evolution of a granular sample submitted to an oedometric test (i.e. R= constant) by increasing the vertical stress $\sigma_v$ from $\sigma_0$. The initial state of the sample will be assumed isotropic and the initial stress too (i.e. $\sigma_{10} = \sigma_{20} = \sigma_{30} = \sigma_0$). At last, we will consider that principal-stress and principal-strain directions remain parallel to one another all along the test due to the symmetry of the system.

Due to the symmetry of the initial state (isotropic stress, isotropic material), the incremental response of this state shall be characterised by two zones (one for $||\delta\sigma|| > 0$, the other for $||\delta\sigma|| < 0$ ), each one is characterised by two independent parameters , i.e. a pseudo Young modulus $C_0$ and a pseudo Poisson coefficient $\nu$. So, in a given zone one can write:

$$
\begin{pmatrix} de_1 \\ de_2 \\ de_3 \end{pmatrix} = \text{-}C_o \begin{pmatrix} 1 & -n & -n \\ -n & 1 & -n \\ -n & -n & 1 \end{pmatrix} \begin{pmatrix} ds_1 \\ ds_2 \\ ds_3 \end{pmatrix}
\tag{1}
$$

As paths which are considered here concern only those ones with a mean stress increase, there is only one set of pseudo Young modulus and pseudo Poisson coefficient of interest. This set can be determined from triaxial test curves performed at $\sigma_2 = \sigma_3 = $cste, since the slope at the origin of the curve $\sigma_1$ vs. $\varepsilon_1$ is just $1/C_o$ and since $\nu$ is related to the dilatancy $K = \text{-}\delta\varepsilon_v/\delta\varepsilon_1 = \text{-}(\partial\varepsilon_v/\partial\varepsilon_1)_{\sigma2 = \sigma3 = \text{cste}}$ at the origin for a test performed at $\sigma_2 = \sigma_3 = $cste by:

$$K = 2\nu\text{-}1 \tag{2}$$

In the same way, one can determine the stress increment $\delta\sigma_2 = \delta\sigma_3$ corresponding to an increment of stress $\delta\sigma_1$ in the case of an oedometric stress by imposing $\delta\varepsilon_2 = \delta\varepsilon_3 = 0$ in Eq. (1); this leads to:

$$\delta\sigma_2/ \ \delta\sigma_1 = \delta\sigma_3/\delta\sigma_1 = (\partial\sigma_2/\partial\sigma_1)_{\text{oedometric}} = \nu/(1\text{-}\nu) \tag{3}$$

When stress increases, deformation increases and incremental law varies so that the Eq. (1) above does not remain constant. Furthermore, according to the symmetry of the oedometric test, one expects the incremental law to depend no more on two parameters, but on four constants $C_0, \nu, \nu', \nu''$:

(4)

$$\begin{pmatrix} d\mathbf{e}_1 \\ d\mathbf{e}_2 \\ d\mathbf{e}_3 \end{pmatrix} = -C_o \begin{pmatrix} 1 & -\mathbf{n}' & -\mathbf{n}' \\ -\mathbf{n} & \mathbf{a} & -\mathbf{n}'' \\ -\mathbf{n} & -\mathbf{n}'' & \mathbf{a} \end{pmatrix} \begin{pmatrix} d\mathbf{s}_1 \\ d\mathbf{s}_2 \\ d\mathbf{s}_3 \end{pmatrix}$$

One can demonstrate that $\nu'=\nu$ since the work shall not depend on the details of the incremental stress path, but shall only depend on the total one $\underline{\delta\underline{\sigma}}$. For sake of simplicity, we will consider in a first step that the material response remains isotropic so that the incremental relation Eq. (1) remains depending on two parameters only, i.e. $C_o$ and $\nu$. These two parameters can be determined directly from experimental triaxial curves at $\sigma_2=\sigma_3=c^{ste}$: $C_o$ remains the inverse of the tangent to the curve ($\sigma_1$ vs. $\varepsilon_1$) at a given $\sigma_1$, and $\nu$ is related to the slope $-K$ of the curve ($\varepsilon_v$ vs. $\varepsilon_1$) at a given $\varepsilon_1$ by :

$$C_o= -1/(\partial\sigma_1/\partial\varepsilon_1)_{\sigma2=cste} \tag{5a}$$

$$\nu=(1+K)/2 \qquad\qquad \text{with} \quad K=-(\partial\varepsilon_v/\partial\varepsilon_1)_{\sigma2=cste} \tag{5b}$$

Hence, the evolution of the stress in an oedometer is given by Eq. (3) where $\nu$ varies. As a matter of fact, the way $\nu$ and K varies with $\sigma_1$ is a well known experimental result of soil mechanics, which is called the Rowe relation, (see above). This one writes:

$$\sigma_1/\sigma_3 = (1+K)\tan^2(\pi/4+\phi/2) = (1+K)(1+\sin\phi)/[1-\sin\phi]= (1+M)(1+K) \tag{6a}$$

$$\text{or} \qquad\qquad q\text{-}K\ \sigma_3=M\ \sigma_3\ (1+K) \tag{6b}$$

with $q = (\sigma_1 - \sigma_3)$ , $M= 2\sin\phi/(1-\sin\phi)$ and $\phi$ the friction angle in the critical state .

Let us recall an other time that Eq. (6) has been established by Rowe in the context of the plasticity theory and not in the one of elasto-plasticity and of the incremental modelling so that its use here may surprise. Nevertheless, Eq. (6) does fit experimental data at $\sigma_2= \sigma_3=c^{ste}$ so that it shall be used here too so that Eq. (1) fits both experimental results at $\sigma_2= \sigma_3=c^{ste}$ and at $p=c^{ste}$ ; but we will discuss this point later. At the moment, we will use this relation to close the problem. So, the evolution of the stress field in an oedometer test performed by increasing continuously $\sigma_1$ is obtained by combining Eqs. (3), (5) and (6). This leads to the differential equation:

$$\delta\sigma_2/\ \delta\sigma_1 = \ \delta\sigma_3/\delta\sigma_1 \ =\nu/(1\text{-}\nu)=(1+K)/(1\text{-}K)= \ \{\sigma_1/[\sigma_3\ \ (1+M)]\}/\{[\sigma_1\text{-}2\sigma_3(1+M)]/[\sigma_3\ (1+M)]\}$$

or:

$$\partial\sigma_3/\partial\sigma_1 = (\sigma_1/\sigma_3)/\{[2(1+M)-(\sigma_1/\sigma_3)]\} = (\sigma_3/\sigma_1)/\{2(1+M)(\sigma_3/\sigma_1)^2-(\sigma_3/\sigma_1)\} \qquad (7)$$

So, starting from a given ratio of $(\sigma_3/\sigma_1)$ the stress field evolves according to Eq. (7). Let us first demonstrate the existence of a stress ratio $(\sigma_1/\sigma_3)_{oed}$ such as $\partial\sigma_3/\partial\sigma_1 = \sigma_3/\sigma_1$ ; we will then demonstrate it is the stable working point toward which the system shall evolve under an oedometer test. Imposing $\partial\sigma_3/\partial\sigma_1 = \sigma_3/\sigma_1$ in Eq. (7) leads to $2(1+M)(\sigma_3/\sigma_1)^2-(\sigma_3/\sigma_1)=1$ which has a unique positive solution:

$$(\sigma_3/\sigma_1)_{oed} = [1+(9+8M)^{1/2}]/[4(1+M)] \qquad (8)$$

$$= (1/4)[(1-\sin\phi)/(1+\sin\phi)][1+\{[9+7\sin\phi]/[1-\sin\phi]\}^{1/2}] \qquad (8)$$

Combining Eqs. (3) & (5b), one gets $\partial\sigma_3/\partial\sigma_1=(1+K)/(1-K)$ ; remembering that experimental triaxial curves at constant $\sigma_2=\sigma_3$ exhibit a continuous increase of K in the range $1<\sigma_1/\sigma_3<M+1$, which is the range of interest, it is straight forward to show that $\partial\sigma_3/\partial\sigma_1 >1$ if $(\sigma_3/\sigma_1)< (\sigma_3/\sigma_1)_{oed}$ , and that $\partial\sigma_3/\partial\sigma_1 <1$ in the contrary , i.e. $(\sigma_3/\sigma_1)> (\sigma_3/\sigma_1)_{oed}$ ; so, any misfit between the real value $(\sigma_3/\sigma_1)$ and $(\sigma_3/\sigma_1)_{oed}$ will be reduced when increasing the load $\sigma_1$. In conclusion, $(\sigma_3/\sigma_1)_{oed}$ is an attracting value which shall be obtained for large enough $\sigma_1$.

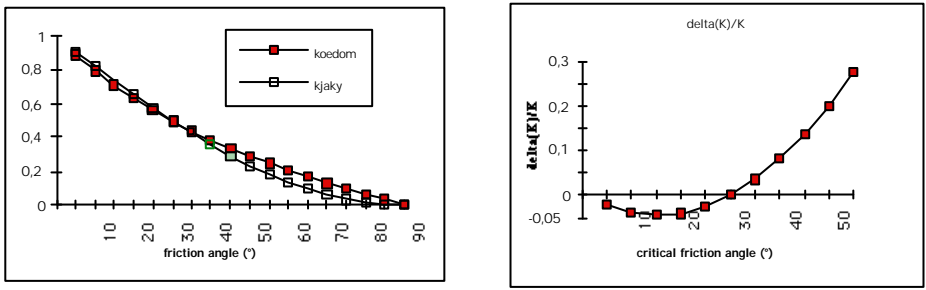

**Figure 1:** Comparison between the oedometric modulus $k_{oedometric}$ as calculated from triaxial test result, i.e. $k_{oedometric}= \sigma_3/\sigma_1 = [1+(8M+9)^{1/2}]/[4M+4]$ , and the experimental best fit of Jaky, i.e. $k_{jaky}= 1-\sin\phi$, vs. the friction angle $\phi$ of perfect plasticity.  Fig. 1a: the two dependencies. Fig. 1b: their relative difference.

This attracting value corresponds to a single value of the dilatancy since it is such as it satisfies at the same time the Rowe relation $(\sigma_1/\sigma_3=[1+M][1+K])$ and the oedometric stress ratio $(\sigma_1/\sigma_3=[1-K]/[1+K])$ of pseudo elastic material having the Poisson coefficient given by $\nu=(1+K)/2$. This leads to a unique possibility which is (Evesque 1997):

$$K_{oedometric}=[(8M+9)^{1/2}-2M-3]/(2M+2) \qquad (9)$$

This approach predicts the ratio $(\sigma_3/\sigma_1)_{oed}$ of principal stresses; it can be compared to the experimental values, a good fit of which is know as the Jaky law (Jaky 1944 & 1948) and depends on the friction angle $\phi$:

$$(\sigma_3/\sigma_1)_{oed\ experimental} = (\sigma_3/\sigma_1)_{Jaky} = 1-\sin\phi \qquad (10)$$

We compare these two values as a function of $\phi$ in Fig. 1a, and their difference in Fig. 1b ; the agreement is correct.

### *Discussion and Conclusion*

Due to symmetry considerations on the characteristics of an oedometric stress, it has been established that the most general incremental behaviour shall be written as in Eq. (4) in which $\nu=\nu'$. In this case, oedometric test is characterised by

$$(\partial\sigma_3/\partial\sigma_1)_{oedometric}= (\partial\sigma_2/\partial\sigma_1)_{oedometric} = \nu/(\alpha-\nu") \qquad (11)$$

In Eq. (11), $\nu$ is still related to the dilatancy K of a triaxial test at $\sigma_2=\sigma_3=$cste by Eq. (5b) , i.e. $K=2\nu-1$. So, Eq. (11) demonstrates that both approaches lead to the same prediction if

$$1-\nu=\alpha-\nu" \qquad (12)$$

Let us discuss this point now. In principle, the two parameters $(\alpha,\nu")$ of Eq. (4) can be determined from experiment but I have not found such a study in the literature. Anyway, one can try and circumpass this difficulty and remark that the starting state is totally isotropic (isotropic state, isotropic stress); it is described by Eq. (1) so that $\nu=\nu'=\nu"$ and $\alpha=1$. As one expects that $\nu"$, $\alpha$ shall evolve slowly from this set of values as far as the deformation remains small because the evolution of these coefficients is related to the change of the contact distribution between the grains.

Indeed, from an experimental point of view, the deformation $\varepsilon_1$ remains small since the $\sigma_1/\sigma_3$ ratio remains smaller than M+1. Furthermore, it is known that an axisymmetric undrained test (i.e. $\delta\varepsilon_v=0$) keeps the pressure $p=(\sigma_1+\sigma_2+\sigma_3)/3$ constant for a large part of the evolution; for instance, for over consolidated granular material this is true till $(\sigma_1-\sigma_3)/\sigma_3$ reaches the value M, that-is-to-say much after having reached the oedometric ratio; this strengthens the hypothesis that the response remains isotropic during a large part of the compression and makes more valid the assumption used.

Unfortunately, the range over which a constant mean pressure is observed during an undrained test is smaller for normally consolidated sand sample and for clay. We demonstrate now that the domain of validity of the proposed approach is just that one for which an undrained test occurs at constant pressure($\delta p=0$) : We remark that condition $\delta\sigma_1+\delta\sigma_2+\delta\sigma_3=\delta p=0$ combined with Eq. (4) imposes:

$$\{(1-2v)\ \delta\sigma_1 + (\alpha-v'-v'')\ (\delta\sigma_2 + \delta\sigma_3)\}=0 \qquad (13)$$

which implies

$$\delta p=0 \qquad => \qquad (1-v) = (\alpha-v'') \qquad (14)$$

since $v'=v$. Eq. (14) is just the condition required for the above approach (see Eq. 12). So, this implies in turn that the oedometric path $(\partial\sigma_3/\partial\sigma_1)_{oedometric} = (\partial\sigma_2/\partial\sigma_1)_{oedometric} = v/(\alpha-v'')$ can be written as $(\partial\sigma_3/\partial\sigma_1)_{oedometric} = v/(1-v)$.

***In conclusion***, this approach is based on an incremental modelling which postulates that the deformation of a soil or of a granular material due to a stress increment is linear in a zone which contains the triaxial tests both at constant mean pressure and at $\sigma_2=\sigma_3=c^{ste}$ ; it allows to predict the oedometric stress ratio $k_0 =(\sigma_1/\sigma_3)_{oed}$ tends to a constant when increasing $\sigma_1$ continuously and relates this asymptotic value to the friction angle $\varphi$ and to the dilatancy via the Rowe's equation. The validity of the approach is probably due to the fact that the stress ratio $\sigma_1/\sigma_3$ is never reachs the one M of the characteristic state; so the deviatoric strain remains small. As the size of the zone where the response is incrementally linear seems rather large, one may expect the number of plastic mechanisms to be either small or quite large.

The validity of this approach means likely that the incremental formalism describes the soil behaviour in a much simpler way than the pure single elasto-plastic mechanism which is used in Granta-gravel and Cam-clay modelling. It means also that the derivation proposed by Rowe of the Rowe's equation may be rather fortuitous since it is based on a single plastic mechanism which cannot be written in the way of Eq. (1) and whose elasto-plastic generalisation would change the meaning of the dilatancy mechanism. To strengthen this idea, it is worth remarking that the derivation proposed by Rowe does not respect the symmetry rules a deformation of an isotropic sample submitted to an isotropic stress shall obey since it does not predict the same dilatancy whatever the stress increment direction. This discussion may help understanding why the discrepancy between the experimental dilatancy measured under an isotropic stress and the one predicted by Rowe's equation is larger in the vicinity of the isotropic stress than further. However, this incremental approach cannot help explaining the existence of the Rowe's formula. Instead, the present approach tries and uses a unified scheme to show the existence of a link between different well established experimental facts (Rowe's equation, Jaky's law, undrained results), so that it demonstrates the coherence of these experimental knowledges.

The electronic arXiv.org version of this paper has been settled during a stay at the Kavli Institute of Theoretical Physics of the University of California at Santa Barbara (KITP-UCSB), in june 2005, supported in part by the National Science Fundation under Grant n° PHY99-07949.

*Poudres & Grains* can be found at :
http://www.mssmat.ecp.fr/rubrique.php3?id_rubrique=402